\documentclass[sigconf]{acmart}
\usepackage{microtype}
\usepackage{graphicx}
\usepackage{subfigure}
\usepackage{balance}
\usepackage{booktabs} % for professional tables
\usepackage{amsthm}
\usepackage{array}
\usepackage{graphicx}
\usepackage{clrscode}
\usepackage{subfigure}
\usepackage{multirow}
\usepackage{multicol}
\usepackage{float}
\usepackage{color}
\usepackage{xcolor}
\usepackage{amsopn}
\usepackage{mathrsfs}
\usepackage{mathtools}
\usepackage{amsmath}
\usepackage{booktabs}
\usepackage{arydshln}
\usepackage{hyperref}
\usepackage{blkarray}
\usepackage{enumerate}
\usepackage{courier}
\usepackage{mathrsfs}
\usepackage{rotating}
\usepackage{bm}
\usepackage{wrapfig}
\usepackage{subfigure}
\usepackage{array}
\usepackage{ragged2e}
\usepackage{hyperref}
\usepackage{amsmath}
\usepackage{threeparttable}

\usepackage{adjustbox} % For adjusting table size
\usepackage{enumitem}

\usepackage{comment}

\usepackage[ruled,linesnumbered]{algorithm2e}

\SetKwComment{comment}{ $triangleright$ \ }{}

\theoremstyle{definition}
\newtheorem{definition}{Definition}
\theoremstyle{theorem}
\newtheorem{mytheory}{Theorem}

\theoremstyle{proof}

\theoremstyle{remark}
\newtheorem*{remark}{Remark}

\AtBeginDocument{%
  \providecommand\BibTeX{{%
    \normalfont B\kern-0.5em{\scshape i\kern-0.25em b}\kern-0.8em\TeX}}}

\copyrightyear{2021} 
\acmYear{2021} 
\setcopyright{acmcopyright}
\acmConference[CIKM '21] {Proceedings of the 30th ACM International Conference on Information and Knowledge Management}{November 1--5, 2021}{Virtual Event, Australia.}
\acmBooktitle{Proceedings of the 30th ACM Int'l Conf. on Information and Knowledge Management (CIKM '21), November 1--5, 2021, Virtual Event, Australia}
\acmPrice{15.00}
\acmISBN{978-1-4503-8446-9/21/11} 
\acmDOI{10.1145/3459637.3482305}
\begin{document}
\author{Mengyue Yang$^{1,\#}$, Quanyu Dai$^{2,\#}$, Zhenhua Dong$^{2}$, Xu Chen$^{3,4,*}$, Xiuqiang He$^{2}$, Jun Wang$^{1}$}\thanks{$\#$ Equal Contribution. * Corresponding Author}
\affiliation{
\institution{$^1$ University College London $^2$ Noah's Ark Lab, Huawei $^3$ Beijing Key Laboratory of Big Data Management and Analysis Methods $^4$Gaoling School of Artificial Intelligence Renmin University of China}
\city{}
\country{}
} 
% \affiliation{
% \institution{$^3$School of Software, Tsinghua University, $^4$Baidu Incorporation,$^5$Ant Group}
% \city{}
% \country{}
% }
\title{Top-N Recommendation with Counterfactual \\User Preference Simulation}
\begin{abstract}
Top-N recommendation, which aims to learn user ranking-based preference, has long been a fundamental problem in a wide range of applications.
Traditional models usually motivate themselves by designing complex or tailored architectures based on different assumptions.
However, the training data of recommender system can be extremely sparse and imbalanced, which poses great challenges for boosting the recommendation performance.
To alleviate this problem, in this paper, we propose to reformulate the recommendation task within the causal inference framework, which enables us to counterfactually simulate user ranking-based preferences to handle the data scarce problem.
The core of our model lies in the counterfactual question: ``what would be the user's decision if the recommended items had been different?''.
To answer this question, we firstly formulate the recommendation process with a series of structural equation models (SEMs), whose parameters are optimized based on the observed data.
Then, we actively indicate many recommendation lists (called intervention in the causal inference terminology) which are not recorded in the dataset, and simulate user feedback according to the learned SEMs for generating new training samples.
Instead of randomly intervening on the recommendation list, we design a learning-based method to discover more informative training samples.
Considering that the learned SEMs can be not perfect, we, at last, theoretically analyze the relation between the number of generated samples and the model prediction error, based on which a heuristic method is designed to control the negative effect brought by the prediction error.
Extensive experiments are conducted based on both synthetic and real-world datasets to demonstrate the effectiveness of our framework. 
\end{abstract}
\begin{CCSXML}
<ccs2012>
 <concept>
  <concept_id>10010520.10010553.10010562</concept_id>
  <concept_desc>Information systems~Recommender systems</concept_desc>
  <concept_significance>500</concept_significance>
 </concept>
</ccs2012>
\end{CCSXML}

\ccsdesc[500]{Information systems~Recommender systems}

\keywords{Recommender Systems, Bayesian Personalized Ranking, Structure Causal Model, Counterfactuals}
\maketitle
\fancyhead{}

\vspace{-2mm}
\section{Introduction}
Recommendation system basically aims to match a user with her most favorite items.
In a typical recommendation process, the system firstly recommends an item list to a user, and then the user provides feedback on the recommendations.
In real-world scenarios, people only access a small amount of items, which makes the observed dataset extremely sparse.
Recommendation task is usually formulated as a ranking problem.
Early models mostly base themselves on the simple matrix factorization method~\cite{rendle2009bpr}.
In order to achieve better performance and adapt different scenarios, recent years have witnessed much effort on neuralizing the recommendation models~\cite{he2017neural, DBLP:journals/corr/abs-2011-01731}.
However, the effectiveness of neural models usually depend on a large amount of training samples, which contradicts with the aforementioned sparse user behaviors.

\begin{figure}[t]
\centering
\setlength{\fboxrule}{0.pt}
\setlength{\fboxsep}{0.pt}
\fbox{
\includegraphics[width=.95\linewidth]{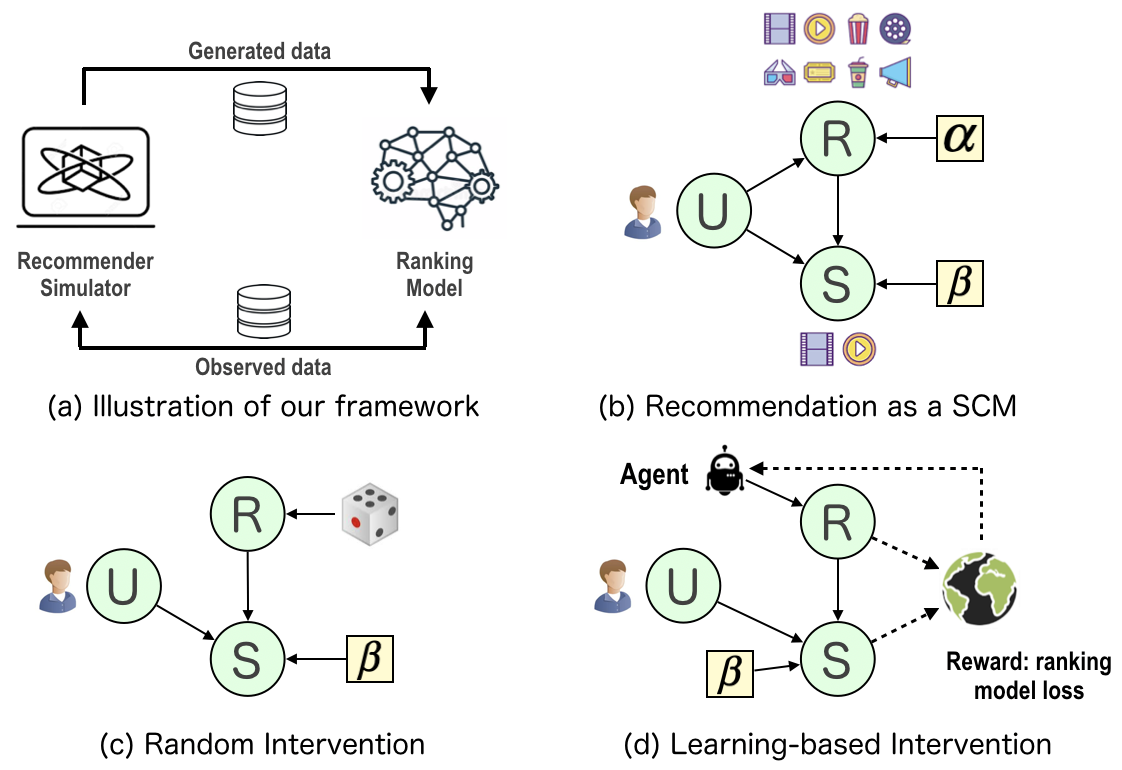}
}
\vspace{-0.cm}
\caption{(a) is the general framework of our idea.
(b) is the structure causal model of the recommender simulator.
(c) and (d) are two different intervention methods.}
\label{intro}
\vspace{-0.2cm}
\end{figure}

In another research line, causal inference (CI) has been recently introduced into the machine learning community to augment the training data for more comprehensive model optimization~\cite{acharyya2020fairyted}.
The basic idea is firstly assuming an underlying structure causal model (SCM), and then learning the model parameters based on the observed data.
At last, the new training samples are generated by actively changing the input variables (called intervention) and collecting the cared outputs.
Such sample enrichment method has been successfully applied to the fields of neural language processing (NLP)~\cite{zmigrod2019counterfactual} and computer vision (CV)~\cite{fu2020counterfactual,chen2019counterfactual,ashual2019specifying}.
In this paper, we adapt this method to the recommendation domain, which is expected to alleviate the contradiction between the more and more heavier neural recommender architectures and the sparse user behaviors.
In a nutshell, the main building block of our idea lies in the counterfactual question: ``what would be the user's feedback if the recommendation list had been different?''.
More specifically, we formulate the recommendation task by a causal graph including three nodes (see Figure~\ref{intro}(b)): \textbf{U} represents the user profiles, \textbf{R} is the recommendation list, and \textbf{S} indicates the selection of the user from \textbf{R}.
The new training samples are generated by collecting the user feedback \textbf{S} with different interventions on \textbf{R}.

While the counterfactual idea seems to be promising, there are many challenges when applying it to the recommendation domain:
to begin with, how to formally define the recommendation task by a structure causal model is still unclear.
Then, the space of the candidate recommendation lists (\textbf{R}) can be very large, and the samples induced from different recommendation lists (i.e., intervention) may vary on the effects of optimizing the target model~\cite{goyal2019counterfactual,wang2017irgan}.
How to design an effective method to select \textbf{R} remains to be an open problem. 
At last, the predefined structural equation models can be not perfect.
How the prediction error influences the quality of the generated samples and how to lower the negative impact need to be carefully considered. 
For solving these challenges, in this paper, we propose a novel \underline{c}ounterfactual \underline{p}ersonalized \underline{r}anking framework (called \textbf{CPR}).
In general, our framework is composed of two parts (see Figure~\ref{intro}(a)), i.e., the target ranking model and the recommender simulator.  
The ranking model is leveraged to provide the final recommendation list, and the recommender simulator aims to assist the optimization of the ranking model by generating additional training samples.
When building the recommender simulator, we follow Pearl's~\cite{pearl2009causality} counterfactual framework to define the recommendation process, where the structural equation models (SEMs) between \textbf{U}, \textbf{R} and \textbf{S} are defined in a stochastic manner and learned by variational inference to capture the randomness in the recommender system.
In order to handle the extremely large space of the recommendation list (i.e., \textbf{R}), we design a learning-based method to select \textbf{R} which can induce more informative training samples. 
Considering that the learned SEMs can be not perfect, we theoretically analyze the relation between its prediction error and the number of generated samples.
Inspired by this theory, we propose a simple but effective strategy to control the quality of the generated samples.

The main contributions of this paper are summarized as follows:
(1) We formulate the recommendation problem within Pearl's causal inference framework, which allows us to generate more training samples for more sufficient model optimization.
(2) We design a learning-based intervention method, which can lead to more informative training samples for optimizing the target ranking model.
(3) We theoretically analyze the relation between the potential prediction error of the structural equation models and the number of generated samples. 
(4) Inspired by the above theory, we propose a heuristic method to control the quality of the generated samples.
(5) We conduct extensive experiments based on both synthetic and real-world datasets to verify the effectiveness of our model.

\vspace{-0.cm}
\section{Preliminaries}
In this section, we firstly recapitulate the key concepts and methodologies in Pearl's causal inference framework.
Then, we briefly introduce the ranking-based recommender models.

\subsection{Pearl's Causal Inference Framework}
Pearl's causal inference framework holds the promise of providing a complete and self-contained tool for studying causalities under both experimental and observational settings~\cite{peters2017elements}.
It is famous for the proposed three layer causal hierarchy, i.e., association, intervention and counterfactual.
In Pearl's causal inference framework, the first key concept is the structure causal model (SCM), which helps to formulate real-world problems with causal languages.

\begin{definition}[Structure Causal Model~\cite{pearl2009causality}]
A structural causal model $\bm{M}$ is composed of a causal graph $\bm{G}$ and a set of structure equation models (SEMs) $\bm{F}$.
$\bm{G}$ is usually a directed graph, where the edges indicate the causal relations between different variables.
The nodes in $\bm{G}$ are classified into two groups: 
(\romannumeral1) \textbf{exogenous nodes} $\bm{U} = \{u_1, . . . , u_{N_U}\}$, which are independent with each other, and summarize the environment when the data was generated,
and (\romannumeral2) \textbf{endogenous nodes} $\bm{X} = \{x_1, . . . , x_{N_X}\}$, which corresponds the variables we need to model in the problem.
$\bm{F} = \{f_i\}_{i=1}^N$ basically parameterizes the node relations, that is, $x_i = f_i(PA_i, u_i)$, where $PA_i$ is the parent of $x_i$ in $\bm{G}$.
\end{definition}

The structure causal model builds the basis of studying the three layer causal hierarchy.
In this paper, we mainly focus on counterfactual estimation.
In general, this tool aims to predict the outcome if several input variables in the causal graph had been different.
We briefly introduce the estimation process in the following definition:

\begin{definition}[Counterfactual Estimation~\cite{pearl2009causality}]
Suppose we have two variable sets $\bm{Y}, \bm{Z}\subset \bm{X}$, and we use small letters $\bm{y}$ and $\bm{z}$ to denote their instantiations.
The notation $p(\bm{y}^M_{\bm{Z} = \bm{z}}(\bm{z'}))$ defines the distribution of $Y$ if $\bm{Z}$ had been set as $\bm{z}$ in structural causal model $M$ given that $\bm{Z}$ is currently observed as $\bm{z'}$.
This distribution corresponds the counterfactual question ``what would be $Y$ if $\bm{Z}$ had been set as $\bm{z}$ given its current observation $\bm{z'}$?'', which is typically derived according to the following three steps:
(\romannumeral1) \textbf{Abduction:} deriving the posterior of the exogenous variables $p(\bm{U}|\bm{Z} = z')$ based on the prior $p(\bm{U})$ and the observation $\bm{Z} = z'$.
(\romannumeral2) \textbf{Action:} modifying $G$ by removing the edges going into $\bm{Z}$ and set $\bm{Z} = z$ (called intervention) to derive $p(y|\bm{Z} = z, \bm{U})$.
(\romannumeral3) \textbf{Prediction:} computing the distribution $P(y^M_{\bm{Z} = z}(z'))$ by $\int_{\bm{U}} p(y|\bm{Z} = z, \bm{U})p(\bm{U}|\bm{Z} = z')\text{d}\bm{U}$.
\end{definition}

The key motivation of the above ``abduction-action-prediction'' procedure is to make sure that \textit{the new (counterfactual) and observed data enjoy the same generation environment, so that they can be compatible and just like to be produced at the same time}~\cite{peters2017elements}.
Since Pearl's causal inference framework is not the focus of this paper, we refer the readers to~\cite{pearl2009causality} for more details.
Based on the tool of counterfactual estimation, one can generate additional training samples to assist the downstream tasks, where a remaining problem is how to indicate $z$ to obtain effective samples.

\subsection{Ranking-based Recommender Models}
Ranking-based recommender models are powerful tools for solving the Top-N recommendation task, which can be optimized in either pair-wise or point-wise manners.
In the pair-wise method, there are usually three inputs, i.e., a user $u$ and a positive-negative item pair $(i, j)$.
The goal is to maximize the user preference margin between the positive and negative items.
Suppose the user preference is estimated from a function $f(\cdot)$, then the optimization target is:
\begin{equation}
\setlength\abovedisplayskip{5pt}%shrink space
\setlength\belowdisplayskip{5pt}
\begin{aligned}
L_{1}(O) = -\sum_{(u,i,j)\in O} \log\sigma(f(u, i)-f(u, j)),
\end{aligned}
\label{pair}
\end{equation}  
where $\sigma(\cdot)$ is the sigmoid function to avoid trivial solutions.
$O=\{(u,i,j)|i\in\mathcal{I}_u^+, j\in\mathcal{I}\setminus\mathcal{I}_u^+\}$ is the set of training samples.
$\mathcal{I}$ is the whole item set, and $\mathcal{I}_u^+$ indicates the set of items which have received $u$'s positive feedback.
For the point-wise method, the input is a user-item pair $(u, i)$, and the user preference estimation is treated as a classification problem,  where $f$ is optimized by the  following cross entropy objective:
\begin{equation}
\begin{aligned}
L_{2}(O) = -\sum_{(u,i)\in O^+}\log f(u,i) -\sum_{(u,i)\in O^-} (1-\log(f(u,i))),
\end{aligned}
\label{point}
\end{equation} 
where $O^+ = \{(u,i)|i\in \mathcal{I}_u^+\}$ and $O^- = \{(u,i)|i\in \mathcal{I}\setminus\mathcal{I}_u^+\}$ are the sets of positive and negative samples.
$O=O^+ \cup O^-$ denotes the complete training set.
Ranking-based recommender models differentiate themselves by the various implementation of $f$.
Simple $f$ can be realized by matrix factorization~\cite{rendle2009bpr}, and more advanced $f$ includes multi-layer perception~\cite{he2017neural,niu2018neural}, attentive neural network~\cite{he2018adversarial} and graph convolutional network~\cite{LightGCN_0001DWLZ020}, among others.

\section{The CPR Framework}
Our framework is composed of two parts (see Figure~\ref{intro}(a)): 
one is the recommender simulator, which is responsible for generating new training samples.
The other is the target ranking model, which is learned based on both of the observed and generated data and leveraged to provide the final recommendation list.
Our framework can be applied to any ranking-based recommender models, and we detail the recommender simulator in the following sections.

\subsection{Recommender Simulator}
To begin with, we reformulate the recommendation problem by a structure causal model $\bm{M} = \{\bm{G}, \bm{F}\}$.
In specific, the causal graph $\bm{G}$ is defined as follows (see Figure~\ref{intro}(b)):
(1) \textbf{U}, \textbf{R} and \textbf{S} are the nodes representing the user, the recommendation list and the positive items\footnote{In practice, the positive feedback can be click, purchase, download, and etc.} selected by the user.
(2) $\textbf{U} \rightarrow \textbf{R}$ encodes the fact that the recommendation list is generated according to the user preference.
(3) $\textbf{U} \rightarrow \textbf{S}$ and $\textbf{R} \rightarrow \textbf{S}$ indicate that the positive items are jointly determined by the recommendation list and user preference.

The structure equation models $\bm{F}$ are defined in a stochastic manner as follows:
\begin{equation}
\setlength\abovedisplayskip{1pt}%shrink space
\setlength\belowdisplayskip{5pt}
\bm{F}:\left\{
\begin{aligned}
\bm{R} &\sim p_{\bm{R}}(\bm{R}|\bm{U}, \bm{\alpha})\\
\bm{S} &\sim p_{\bm{S}}(\bm{S}|\bm{U}, \bm{R}, \bm{\beta}) \\
~~\bm{\alpha},~\bm{\beta} &\sim \mathcal{N}(\bm{0}, \bm{I})
\end{aligned}
\right.
\label{fsem}
\end{equation}
where $\bm{U}$, $\bm{R}$ and $\bm{S}$ are endogenous variables.
$\bm{\alpha}\in \mathbb{R}^{|\mathcal{I}|}$ and $\bm{\beta}\in \mathbb{R}^K$ are exogenous variables.
$p_{\bm{R}}$ is the probability of recommending item list $\bm{R}$.
$p_{\bm{S}}$ is the probability of selecting $\bm{S}$ as the positive feedback. 
$\mathcal{N}(\bm{0}, \bm{I})$ is the standard Gaussian distribution.

\begin{remark}
(1) In order to consider the potential noisy information and randomness in the recommender system, the structure equation models (i.e., $\bm{F}$) are defined in a stochastic manner, which helps to learn user preference in a more accurate and robust manner.
(2) The exogenous variables in the recommendation problem can be explained as the conditions (e.g., system status, user habit, etc.) which induces the currently observed data.
Take movie recommendation as an example, the users may be more likely to watch movies in their leisure time.
So if the data is collected at night, then the movie click probability should be higher.
Such temporal condition is latent but can influence the data generation process, which is expected to be captured by the exogenous variables, and recovered by inferring the corresponding posteriors.
\end{remark}

\subsubsection{Specification of $\bm{F}$}
We implement $\bm{F}$ by considering the unique characters of the recommender system.
Recall that $p_{\bm{R}}$ is the probability of recommending item list $\bm{R}$.
Suppose we have $|\mathcal{I}|$ items in the system, then the number\footnote{Here, if we consider the order of the recommended item, the candidate item space can be even larger.} of candidate item lists is $C_{|\mathcal{I}|}^{|\bm{R}|}$, which is extremely large in the recommendation task.
To alleviate this problem, we ignore the combinatorial effect between different items, and the probability of recommending an item list can be factorized into the multiplication of the single item recommendation probabilities, that is:
\begin{equation}
\setlength\abovedisplayskip{1pt}%shrink space
\setlength\belowdisplayskip{5pt}
\begin{aligned}
p_{\bm{R}}(\bm{R} = \bm{r}|\bm{U} = u, \bm{\alpha}) &=  \prod_{k=1}^{K}p({R}_k = {r}_{k}|\bm{U} = u, \bm{\alpha})\\
&=\prod_{k=1}^{K} \frac{\exp{(\bm{P}_{u}^T\bm{Q}_{r_{k}}+{w}_{r_{k}}^R{\alpha}_{r_{k}})}}{\sum_{j=1}^{|\mathcal{I}|}\exp{(\bm{P}_{u}^T\bm{Q}_{j}}+{w}_j^R{\alpha}_{j})}
\end{aligned}
\label{r}
\end{equation} 
where $\bm{r} = \{r_1, r_2, ..., r_K\}$ is the recommendation list.
For the single item recommendation probability, we predict it by a softmax operator.
We use $\bm{P}\in \mathbb{R}^{|\mathcal{U}|\times d_R}$ and $\bm{Q}\in \mathbb{R}^{|\mathcal{I}|\times d_R}$ to represent the embeddings of the users and items.
They can be initialized by the ID or profile information of the users and items.
$d_R$ is the embedding size, and $\bm{w}^R\in \mathbb{R}^{|\mathcal{I}|}$ is a weighting parameter.

$p_{\bm{S}}$ represents the probability of selecting item set $\bm{S}$ from $\bm{R}$, which is specified as:
\begin{equation}
\setlength\abovedisplayskip{5pt}%shrink space
\setlength\belowdisplayskip{5pt}
\begin{aligned}
&p_{\bm{S}}(\bm{S} = \bm{s}|U = u, \bm{R} = \bm{r}, \bm{\beta}) = \prod_{t=1}^{M} \frac{\exp{(\bm{X}_{u}^T\bm{Y}_{s_{t}}+{w}_{{t}}^S{\beta}_{{t}})}}{\sum_{j=1}^{K}\exp{(\bm{X}_{u}^T\bm{Y}_{r_{j}}}+{w}_{{j}}^S{\beta}_{{j}})}\\
\end{aligned}
\label{s}
\end{equation}
where $\bm{s} = \{s_1, s_2, ..., s_M\}$ is the set of selected items.
$\bm{w}^S\in \mathbb{R}^K$ is a weighting parameter.
The user and item properties are encoded by the metrics of $\bm{X}\in \mathbb{R}^{|\mathcal{U}|\times d_S}$ and $\bm{Y}\in \mathbb{R}^{|\mathcal{I}|\times d_S}$, respectively, and $d_S$ is the embedding size. 
In $p_{\bm{R}}$ and $p_{\bm{S}}$, we represent the users and items with different embedding metrics.
This can more flexibly characterize the system impression model (i.e., $\textbf{U} \rightarrow \textbf{R}$ in Figure~\ref{intro}(b)) and the user response model (i.e., $\textbf{U} \rightarrow \textbf{S}$ in Figure~\ref{intro}(b)), which usually hold different promises in real recommender systems.

\subsubsection{Learning $F$.}
Suppose the training set is $\mathcal{O} = \{(u_i, \bm{r}_i, \bm{s}_i)\}_{i=1}^N$, where $u_i$ is the target user, $\bm{r}_i$ is the recommendation list, $\bm{s}_i$ is the item set selected by $u_i$ from $\bm{r}_i$, N is the sample number.  
For $p_{\bm{R}}$, since the number of items (i.e., $|\mathcal{I}|$) can be very large, it is hard to directly optimize the softmax term in equation~(\ref{r}).
Thus we resort to the negative sampling technique~\cite{rendle2009bpr,he2017neural}, which induces the following optimization target:
\begin{equation}
\setlength\abovedisplayskip{5pt}%shrink space
\setlength\belowdisplayskip{5pt}
\begin{aligned}
&L_R \!= \! \! \!\sum_{i=1}^N \!\sum_{k\in \bm{r}_i} \!\log{(\sigma(\bm{P}_{u_i}^T\bm{Q}_k\!+ \!w_k^R\!{\alpha}_{k}))}\! + \!\!\!\!\!\sum_{k^-\in \bm{r}^-_i}\! \!\!\log{(1\!-\!\sigma(\bm{P}_{u_i}^T\bm{Q}_{k^-}\!+ \!w_{k^-}^R\!{\alpha}_{k^-}))} 
\end{aligned}
\label{r-o}
\end{equation}
where 
$\{\bm{P}, \bm{Q}, \bm{w}^R\}$ are the parameters to be learned. $\bm{r}^-_i$ is the set of negative samples, which are randomly sampled from the non-recommended items.
$\bm{\alpha}=[\alpha_i]$ is sampled\footnote{For improving the accuracy, $\bm{\alpha}$ is sampled multiple times for implementation, and the loss is averaged across different $\bm{\alpha}$'s.} from $\mathcal{N}(\bm{0}, \bm{I})$. 
With this objective, the parameters are optimized to maximize the recommended item probability and simultaneously minimize the ones which are not presented to the users.

For $p_{\bm{S}}$, since the length of the recommendation list (i.e., K) is usually not large, we directly maximize the following livelihood without approximation:
\begin{equation}
\setlength\abovedisplayskip{5pt}%shrink space
\setlength\belowdisplayskip{5pt}
\begin{aligned}
L_S = \sum_{i=1}^N\sum_{t=1}^{M} \log(\frac{\exp{(\bm{X}_{u_i}^T\bm{Y}_{s_{it}}+\bm{w}_{{t}}^S{\beta}_{{t}})}}{\sum_{j=1}^{K}\exp{(\bm{X}_{u_i}^T\bm{Y}_{r_{ij}}+\bm{w}_{{j}}^S{\beta}_{{j}})}})\\
\end{aligned}
\label{s-o}
\end{equation}
where $\{\bm{X}, \bm{Y}, \bm{w}^S\}$ are the parameters to be optimized. 
$s_{it}$ and $r_{ij}$ are the $t$th and $j$th items in $\bm{s}_i$ and $\bm{r}_i$, respectively. $\bm{\beta}=[\beta_i]$ is sampled from $\mathcal{N}(\bm{0}, \bm{I})$.
It seems that this objective only maximizes the probability of the items which are selected by the user, but with the softmax operator, the probabilities of the non-selected items are automatically lowered.

\subsubsection{Counterfactual estimation based on $F$}
Once we have learned $p_{\bm{R}}$ and $p_{\bm{S}}$, we follow Pearl's ``abduction-action-prediction'' procedure~\cite{pearl2016causal} to conduct counterfactual estimation.
In the step of abduction, we estimate the posterior of $\bm{\alpha}$ and $\bm{\beta}$ given the observed dataset $\mathcal{O}$.
For $\bm{\beta}$, the posterior can be derived based on the following Bayesian rules:
\begin{equation}
\begin{aligned}
p(\bm{\beta}|\mathcal{O}) &\propto p(\bm{\beta}, \mathcal{O})= p(\bm{\beta})p(\mathcal{O}|\bm{\beta})\\
% \propto &\prod_{s=1}^{|\mathcal{I}|}p({\alpha}_s)\prod_{i=1}^N \prod_{k=1}^{K} \frac{\exp{(\bm{P}_{u_i}^T\bm{Q}_{r_{ik}}+\bm{w}_{r_{ik}}^R{\alpha}_{r_{ik}})}}{\sum_{j=1}^{|\mathcal{I}|}\exp{(\bm{P}_{u_i}^T\bm{Q}_{j}}+\bm{w}_j^R{\alpha}_{j})}\\
\end{aligned}
\label{e-r}
\end{equation}
where $p(\mathcal{O}|\bm{\beta})$ can be easily derived based on equation~(\ref{s}).
Recall that our goal is to sample from the posteriors.
However, the result of equation~(\ref{e-r}) is too complex for sampling, which motivates us to use variational inference~\cite{blei2017variational} for approximation.
In specific, we firstly define a Gaussian distribution $q_{\bm{\phi}}(\bm{\beta})\sim \mathcal{N}(\bm{\mu}, \bm{\sigma})$, where $\bm{\phi} = \{\bm{\mu}, \bm{\sigma}\}$ is set of learnable parameters. 
Then we optimize $\bm{\phi}$ by minimizing the KL-divergence between $q_{\bm{\phi}}(\bm{\beta})$ and $p(\bm{\beta}|\mathcal{O})$, where the evidence lower bound (ELBO)~\cite{blei2017variational} we need to maximize is:
\begin{equation}
\begin{aligned}
\text{ELBO} = {\text{E}_{q_{\bm{\phi}}(\bm{\beta})}[{\log{{p(\bm{\beta},\mathcal{O})}}}]-\text{E}_{q_{\bm{\phi}}(\bm{\beta})}[{\log{{q_{\bm{\phi}}(\bm{\beta})}}}]}\\
\end{aligned}
\label{ELBO-r}
\end{equation}
Similarly, we can learn a variational distribution for $p(\bm{\alpha}|\mathcal{O})$.
Due to the space limitation, readers are referred to ~\cite{blei2017variational} for more technical details of the variational inference.

In the action step, we select a user $\hat{u}$, and set $\bm{R} = \hat{\bm{r}}$.
When making prediction, we firstly sample $\hat{\bm{\beta}}$ from $q_{\bm{\phi}}(\bm{\beta})$, and then compute the probability of item $\hat{r}_k$ by:
\begin{equation}
\setlength\abovedisplayskip{2pt}%shrink space
\setlength\belowdisplayskip{2pt}
\begin{aligned}
&p_{\bm{S}}(\hat{r}_k|U = \hat{u}, \bm{R} = \hat{\bm{r}}, \hat{\bm{\beta}}) = \frac{\exp{(\bm{X}_{\hat{u}}^T\bm{Y}_{\hat{{r}}_k}+\bm{w}_{k}^S\hat{\beta}_{k})}} {\sum_{j=1}^{K}\exp{  (\bm{X}_{\hat{u}}^T \bm{Y}_{\hat{r}_{j}} +\bm{w}_{j}^S \hat{\beta}_j) }}
\end{aligned}
\label{s-infer}
\end{equation}
At last, $\hat{\bm{s}}$ is predicted by collecting M items with the highest probabilities.

\subsection{Learning-based Intervention}
Based on the above designed recommender simulator, one can attempt different $\hat{\bm{r}}$'s and derive the corresponding $\hat{\bm{s}}$'s to form new (counterfactual) training samples.
A nature question is how to set $\hat{\bm{r}}$.
Straightforwardly, one can explore $\hat{\bm{r}}$ in a random manner (e.g., show the users with random recommendation lists).
However, as mentioned before, the space of $\hat{\bm{r}}$ can be extremely large, and it is well known that different training samples are not equally important for model optimization~\cite{wang2017irgan}, thus the random method can be less effective in hitting the optimal samples.
In order to achieve better optimization results, we design a learning-based method to select $\hat{\bm{r}}$.
Our key idea is to make the generated samples more informative for the target ranking model.
It has been studied in the previous work~\cite{gao2018self,fu2020counterfactual} that the samples with larger loss can usually provide more knowledge for the model to learn (i.e., more informative).
They can well challenge the model and bring more inspirations to improve the performance.
Following these studies, we use the loss of the target ranking model as the reward, and build a learning-based method to generate $\hat{\bm{r}}$. 

Formally, suppose the target ranking model is $f$, and we denote its loss function as $L_{f}$, which can be specified as equation~(\ref{pair}) or~(\ref{point}).
The goal of the agent is to conduct actions (generating recommendation lists) which can lead to larger $L_{f}$.
Considering that the action space can be very large, we follow the previous work~\cite{zhao2017deep} to learn a Gaussian policy to generate the continuous item center of $\hat{\bm{r}}$, after which the discrete item IDs are recovered based on equation~(\ref{r}). The final learning objective is: 
\begin{equation}
\setlength\abovedisplayskip{2pt}%shrink space
\setlength\belowdisplayskip{2pt}
\begin{aligned}
L_{\text{Agent}}(\bm{\theta}) &= \mathbb{E}_{\hat{\bm{\tau}}_t\in\pi(\cdot|\hat{u}, \bm{\theta})}[\sum_{t=1}^T L_f(C(\hat{\bm{\tau}}_t))]\\
\end{aligned}
\label{RL}
\end{equation}
where $\pi$ is the Gaussian policy implemented as a two-layer fully connected neural network with ReLU as the activation function.
$\hat{u}$ indicates the target user.
$L_f(C(\hat{\bm{\tau}}_t))$ denotes the loss of the generated training samples $C(\hat{\bm{\tau}}_t)$.
We generate T sets of training samples, and each one is derived from $\hat{\bm{\tau}}_t$ based on the following steps:

\noindent
$\bullet$ Deriving $\hat{\bm{r}}$ by selecting K items near the item center $\hat{\bm{\tau}}_t$ according to the scores $\{(\hat{\bm{\tau}}_t^T\bm{Q}_{{k}}+\bm{w}_{{k}}^R{\alpha}_{{k}})\}_{k=1}^{|\mathcal{I}|}$. 

\noindent
$\bullet$ Deriving $\hat{\bm{s}}$ by selecting M items with the largest probabilities of $p_{\bm{S}}(\hat{r}_k|U = \hat{u}, \bm{R} = \hat{\bm{r}}, \hat{\bm{\beta}})~~(k\in [1,K])$ .
 % $\{(\bm{X}_{\hat{u}}^T\bm{Y}_{{{r}}_l}+\bm{w}_{l}^S{\beta}_{l})\}_{l=1}^{K}$ (i.e., equation~(\ref{s}));

\noindent
$\bullet$ For objective~(\ref{pair}), $C(\hat{\bm{\tau}}_t) = \{(\hat{u},i,j)|i\in \hat{\bm{s}}, j\in \hat{\bm{r}}\setminus\hat{\bm{s}}\}$.
For objective~(\ref{point}), $C(\hat{\bm{\tau}}_t) = \{(\hat{u},i)|i\in \hat{\bm{s}}\} \cup \{(\hat{u},i)|i\in \hat{\bm{r}}\setminus\hat{\bm{s}}  \}$.

We summarize the complete learning process in Algorithm~\ref{alg:Framwork} and~\ref{RL-alg}.
In specific, the target ranking model is firstly trained based on the original dataset.
And then, we generate many counterfactual training samples based on the Gaussian policy.
At last, the target ranking model is retrained based on the generated datasets.

\setlength{\textfloatsep}{0.2cm}
\begin{algorithm}[t] 
\caption{Learning algorithm of CPR} 
\label{alg:Framwork} 
Learn the SEMs based on objective~(\ref{r-o}) and~(\ref{s-o}). \\
Train the target model $f$ based the original dataset.\\
Pre-train the Gaussian policy $\pi$ according to Algorithm~\ref{RL-alg}.\\
\For{Iteration number k in [1,~K]}{ 
   \For{each user $\hat{u}$}{
       Select an action $\hat{\bm{\tau}}_t$ according to $\pi(\hat{\bm{\tau}}_t|\hat{u}, \bm{\theta})$.\\
       Derive $C(\hat{\bm{\tau}}_t)$ based on $\hat{\bm{\tau}}_t$.\\
       Train $f$ based on $C(\hat{\bm{\tau}}_t)$.\\
   }
}
Making recommendation based on the ranking model $f$.\\
\end{algorithm}

\begin{algorithm}[t]
\caption{Learning algorithm of $\pi$} 
\label{RL-alg} 
Initialize the parameter set $\bm{\theta}$ in the Gaussian policy.\\
\For{episode number in [1,~K]}{ 
    \For{t in [1,~T]}{
        Sample a user $\hat{u}$.\\
        Select an action $\hat{\bm{\tau}}_t$ according to $\pi(\hat{\bm{\tau}}_t|\hat{u}, \bm{\theta}) + N_{t}$, where $N_{t}$ is an exploration noise.\\
        Sample $\bm{\alpha}$ and $\bm{\beta}$ from their posteriors.\\
        Derive $C(\hat{\bm{\tau}}_t)$ and the reward $L_f(C(\hat{\bm{\tau}}_t))$ based on $\hat{\bm{\tau}}_t$.\\
    }
    $\bm{\theta} \leftarrow \bm{\theta} + \alpha [\sum_{t=1}^T L_f(C(\hat{\bm{\tau}}_t)) \nabla_{\theta}\log{(\pi(\bm{r}_t|\theta))}]$.\\
    % Update the ranking model $f$ by $C(\hat{\bm{\tau}}_t)$.\\
}
\end{algorithm}

\subsection{Theoretical Analysis}\label{theory}
In this section, we theoretically analyze the proposed framework within the probably approximately correct (PAC) learning framework.
We focus on the pair-wise learning objective, and the conclusions can be easily extended to the point-wise case.

\subsubsection{Theoretical insights on the learning-based intervention method.}
The key motivation of the learning-based method is to generate harder samples, so that the target model can be more informed and achieve better performance.
The effectiveness of this idea has been empirically demonstrated in the previous work~\cite{wang2017irgan,gao2018self}.
Here, we provide a theoretical justification based on the following theory:
\begin{mytheory}
\label{t}
Suppose the users' feedback on an item pair $(i, j)$ is probabilistic, and we can observe $i>j$ and $i<j$ with the probabilities of $\eta$ and $1-\eta$, respectively.
Then, $\eta$ can actually measure the hardness of the sample, when $\eta$ is closer to $\frac{1}{2}$, then the sample is harder, since the propensity between the items is more ambiguous.
Suppose $\epsilon, \delta \in (0,1)$, and we use a simple voting mechanism to determine the relation between $i$ and $j$, then we need to have $\frac{\log{\frac{1}{\delta}}}{2(1-2\eta)^2}$ samples on $(i, j)$ to ensure that the prediction error is smaller that $\delta$.
\end{mytheory}

This theory suggests that we need to generate more samples (i.e., larger $\frac{\log{\frac{1}{\delta}}}{2(1-2\eta)^2}$) for the harder (i.e., smaller $\eta$) item pairs.
It agrees with our goal in learning-based intervention method, where the produced samples are expected to be hard and can well challenge the target model.
We present the proof of this theory in the Appendix.

\subsubsection{Imperfection of SEMs.}
Since the new samples are generated based on the learned SEMs, careful readers may have concerns on how would the approximation error of $\bm{F}$ influence the sample generation.
To shed lights on this problem, we theoretically analyze the relation between the prediction error of $\bm{F}$ and the number of generated samples.
To begin with, we assume that the recommender simulator can recover the true ranking of the item pairs with a noisy parameter $\zeta\in (0,0.5)$, i.e., suppose the true triplet is $(u,i,j)$, then the recommender simulator generates the true (i.e., $(u,i,j)$) and wrong (i.e., $(u,j,i)$) samples with the probabilities of $1-\zeta$ and $\zeta$, respectively. As the special cases, $\zeta  = 0$ means the recommender simulator is perfect, and there is no noisy information in the produced samples.
$\zeta  = 0.5$ means the recommender simulator is a totally random predictor, and the generated data is nothing but noise. 
Then, we have the following theory:
\begin{mytheory}
\label{t1}
Suppose $\epsilon, \delta \in (0,1)$, and $f\in \mathcal{F}$ is the target ranking model.
If the number of training samples is larger than $\frac{2\log{(\frac{2|\mathcal{F}|}{\delta})}}{\epsilon^2(1-2\zeta)^2}$, then the estimation error of $f$ in terms of ranking prediction is smaller than $\epsilon$ with the probability larger than $1-\delta$.
\end{mytheory}
The proof of this theory is similar to that of theory 1 in~\cite{wang2021counterfactual}. From this theory, we can see, as the noisy parameter $\zeta$ becomes larger, more samples (i.e., $\frac{2\log{(\frac{2|\mathcal{F}|}{\delta})}}{\epsilon^2(1-2\zeta)^2}$) are needed to achieve sufficiently well performance (the prediction error is smaller than $\epsilon$).
Extremely, when $\zeta \rightarrow 0.5$, we need to produce infinite training samples. 
This implies that if the recommender simulator is completely random, then we can not expect well performance by training on the generated data, which is aligned with the intuition.

\subsubsection{Controlling the noisy information.}
Inspired by this theory, we propose a simple but effective method to control the noisy information.
The general idea is to remain the samples with higher confidence.
In specific, for a given recommendation list $\hat{\bm{r}}$, we use the selection probability $p_{\bm{S}}(\hat{r}_k|U = \hat{u}, \bm{R} = \hat{\bm{r}}, \hat{\bm{\beta}})$ to measure the confidence of the recommender simulator.
% we argue that the recommender simulator should be more confident on regarding the items with larger (or lower) selection probability $p_{\bm{S}}(\hat{r}_k|U = \hat{u}, \bm{R} = \hat{\bm{r}}, \hat{\bm{\beta}})$ as positive (or negative) samples.
We denote by $\hat{\bm{s}}^k_+$ and $\hat{\bm{s}}^k_-$ the sets of k items with the largest and smallest selection probabilities.
Then the training set for the pair-wise objective is built as $C(\hat{\bm{\tau}}_t) = \{(\hat{u},i,j)|i\in \hat{\bm{s}}^k_+, j\in \hat{\bm{s}}^k_-\}$.
For the point-wise objective, $C(\hat{\bm{\tau}}_t)$ is set as $\{ (\hat{u},i)|i\in \hat{\bm{s}}^k_+\} \cup \{(\hat{u},i)|i\in \hat{\bm{s}}^k_-  \}$.
In this method, if $k$ is small, then the model has more confidence on the generated samples, which may reduce the noisy information, but at the same time, less samples can be generated, which may lead to insufficient model optimization. If $k$ is large, more samples will be generated, but the noisy level can also be high. 
In this sense, $k$ is actually a parameter to trade-off the noisy information and the number of generated samples.

\section{Experiments}
In this section, we conduct extensive experiments to verify the effectiveness of our framework.
In the following, we begin with the experiment setup, and then report and analyze the results.
 
\subsection{Experiment Setup}
\subsubsection{Datasets} 
Our experiments are based on both synthetic and real-world datasets.
With the synthetic dataset, we aim to evaluate our framework in a controlled manner under clean environment.
By the real-world dataset, we try to verify our framework's effectiveness in real-world settings. 

We follow the previous work~\cite{DBLP:conf/kdd/ZouKCC019} to build the synthetic dataset, where we simulate $N_U (=600)$ users and $N_I (=300)$ items.
For each user $i$ (or item $j$), the preferences $\bm{p}_i\in \mathbb{R}^{d}$ (or properties $\bm{q}_j\in \mathbb{R}^{d}$) are generated from a multi-variable Gaussian distribution $\mathcal{N}(\bm{0},\bm{I})$, where $d$ and $\bm{I}$ represent the vector size and unit matrix, respectively.
In order to generate the recommendation list for user $i$, we firstly compute the score of recommending item $j$ based on a neural network: $r_{ij}= 1-\sigma(\bm{a}^T\kappa_1(\kappa_2([\bm{p}_{i}, \bm{q}_{j}]))+{b})$,
% \begin{equation}
% \begin{aligned}
%     r_{ij}=& 1/(1+exp(\bm{1}^{T}\kappa(\text{ReLU}([\bm{p}_{i}, \bm{q}_{j}])))\\
% \end{aligned}
% \end{equation}
where $\bm{a}\in \mathbb{R}^{2d}$ is specified as an all-one vector, and ${b}$ is set as zero.
We follow~\cite{DBLP:conf/kdd/ZouKCC019} to set $\kappa_1(\cdot)$ and $\kappa_2(\cdot)$ as piecewise functions, and $\kappa_1(x) = x - 0.5$ if $x>0$, otherwise $\kappa_1(x) = 0$, $\kappa_2(x) = x$ if $x>0$, otherwise $\kappa_2(x) = 0$.
Given $r_{ij}$, the probability of recommending item $j$ is $\frac{e^{r_{ij}}}{\sum_{k=1}^{N_I}e^{r_{ik}}}$.
For each user, we generate 25 impression lists, each of which is composed of 5 items.

When generating the feedback of a user $i$ on an item $j$, we follow the previous work~\cite{DBLP:conf/kdd/ZouKCC019} to explore both linear and non-linear user response models, that is:
\begin{equation}
\begin{split}
    &x_{1} =  \bm{a}^T\kappa_1(\kappa_2([\bm{p}_{i}, \bm{q}_{j}]))+{b}\\
    &x_{2} =  \bm{a}^T\kappa_3(\kappa_2([-\bm{p}_{i},-\bm{q}_{j}]))+{b}\\
    &s_{ij}^\text{lin} = \mathbb{I}(\sigma(x_{1}) + N_y - 0.5)\\
    &s_{ij}^\text{nonlin} = \mathbb{I}(\sigma(x_{1} + x_{1}\cdot x_{2}) + N_y - 0.5)\\
\end{split}
\end{equation}
where $\kappa_3(x) = x + 0.5$ if $x<0$, otherwise $\kappa_3(x) = 0$. 
$N_y$ is used to set the noisy level of the datasets, and its default value is 0. 
$\mathbb{I}(x)$ is an indicator function, which is 1 if $x>0$, and 0 otherwise.

The real-world experiments are based on the recently released MIND\footnote{https://msnews.github.io/} dataset~\cite{mind_acl_2020}.
This data is collected from the user behavior log of Microsoft News, and we uses the MIND-small dataset for experiments, where we are provided with 156,965 recommendation lists and 234,468 interactions between 50,000 users and 20,288 items.

\begin{table*}[t]
\caption{Performance comparison based on the synthetic dataset. We present the relative improvements of our framework over the target model in the parentheses. }
\vspace{-0.3cm}
\center
\small
\renewcommand\arraystretch{1.1}
\setlength{\tabcolsep}{5.pt}
\begin{threeparttable}  
\scalebox{.95}{
\begin{tabular}{c|cc|cc|cc|cc}
\hline\hline
Feedback Function &\multicolumn{4}{c|}{Linear}                                                                                           &\multicolumn{4}{c}{Non-Linear}\\\hline
User/Item Dimension         &\multicolumn{2}{c|}{d=16}                                  &\multicolumn{2}{c|}{d=32}                                 &\multicolumn{2}{c|}{d=16}                                 &\multicolumn{2}{c}{d=32}\\\hline
Metrics           & \multicolumn{1}{c}{{HR@10}}&\multicolumn{1}{c|}{{NDCG@10}}&\multicolumn{1}{c}{{HR@10}}&\multicolumn{1}{c|}{{NDCG@10}}&\multicolumn{1}{c}{{HR@10}}&\multicolumn{1}{c|}{{NDCG@10}}&\multicolumn{1}{c}{{HR@10}}&\multicolumn{1}{c}{{NDCG@10}}\\\hline

ItemPop        &0.1450&0.0695&0.1219&0.0597&0.3522&0.1849&0.3289&0.1727\\
ItemKNN        &0.1351&0.0632&0.1188&0.0544&0.3477&0.1663&0.3289&0.1626\\
CDAE           &0.1179&0.0550&0.1071&0.0507&0.1869&0.0917&0.1421& 0.0661\\\hline

BPR       & 0.1567                        & 0.0852                            & 0.1388                          & 0.0712                         & 0.3799                        & 0.2162                            & 0.3636                          & 0.2052                            \\
{CPR-BPR-r}       & 0.1632                        & 0.0861                            & 0.1465                          & 0.0788                         & 0.3748                       & 0.2118                            & 0.3615                          & 0.2017                            \\
{CPR-BPR}   & 0.1652 \scriptsize{(+5.4\%)}                        & 0.0937 \scriptsize{(+9.9\%)}                            & 0.1482 \scriptsize{(+6.7\%)}                          & 0.0760 \scriptsize{(+6.7\%)}                           & {0.4024} \scriptsize{(+5.9\%)}                           &{0.2268} \scriptsize{(+4.9\%)}                             & 0.3705 \scriptsize{(+1.8\%)}                           &0.2028 \scriptsize{(-1.1\%)}                             \\\hline

GMF   &0.1498                          & 0.0796                            & 0.1304                          & 0.0646                            & 0.3654                          & 0.1927                          & 0.3487                          & 0.1909 \\
CPR-GMF-r  &0.1573                          & 0.0760                            & 0.1373                          & 0.0640                            & 0.3830                          & 0.2014                          & 0.3606                          & 0.1886 \\
{CPR-GMF}   & 0.1624 \scriptsize{(+8.4\%)}                           & 0.0744 \scriptsize{(-6.5\%)}                             & 0.1344 \scriptsize{(+3.0\%)}                           & 0.0642 \scriptsize{(-0.6\%)}                            & 0.3808 \scriptsize{(+4.2\%)}                           & 0.1944 \scriptsize{(+0.8\%)}                             & 0.3576 \scriptsize{(+2.5\%)}                           & 0.1891 \scriptsize{(-0.9\%)}                             \\\hline

MLP   &0.1421                          & 0.0705                            & 0.1174                          & 0.0582                            & 0.3603                          & 0.1886                          & 0.3286                          & 0.1718 \\
{CPR-MLP-r}   &0.1422                          & 0.0641                            & 0.1197                          & 0.0593                            & 0.3842                          & 0.1949                          & 0.3626                          & 0.1861 \\
{CPR-MLP}   & 0.1592 \scriptsize{(+12.0\%)}                       & 0.0802 \scriptsize{(+13.7\%)}                            & 0.1331 \scriptsize{(+13.3\%)}                          & 0.0608 \scriptsize{(+4.4\%)}                           & 0.3831 \scriptsize{(+6.3\%)}                          & 0.1940 \scriptsize{(+2.8\%)}                            & 0.3668 \scriptsize{(+11.6\%)}                          & 0.1897 \scriptsize{(+10.4\%)}                            \\\hline

NeuMF   & 0.1524                          & 0.0758                            & 0.1203                          & 0.0598                            & 0.3679                         & 0.1917                            & 0.3350                          & 0.1754 \\
{CPR-NeuMF-r}   & 0.1731                          & 0.0937                            & 0.1582                          & 0.0815                            & 0.3915                         & 0.2122                            & 0.3701                          & 0.1971 \\
{CPR-NeuMF}   & {0.1851} \scriptsize{(+21.4\%)}                        &0.0963 \scriptsize{(+27.0\%)}                           & {0.1667} \scriptsize{(+38.5\%)}                          & {0.0835} \scriptsize{(+39.6\%)}                           & 0.4007 \scriptsize{(+8.9\%)}                           & 0.2166 \scriptsize{(+12.9\%)}                             & 0.3724 \scriptsize{(+11.1\%)}                          & 0.2019 \scriptsize{(+15.1\%)}                            \\\hline

LightGCN        & 0.1503                          & 0.0725                            & 0.1262                          & 0.0605                            & 0.3732                          & 0.1915                            & 0.3451                          & 0.1744                            \\
{CPR-LightGCN-r}  &0.1600                          & 0.0834                            & 0.1317                          & 0.0643                            & 0.3759                          & 0.2008                          & 0.3482                          & 0.1817 \\
{CPR-LightGCN}   & 0.1677 \scriptsize{(+11.6\%)}                           & 0.0879 \scriptsize{(+21.2\%)}                             & 0.1551 \scriptsize{(+22.9\%)}                           & 
0.0836 \scriptsize{(+38.2\%)}                            & 0.3829 \scriptsize{(+2.5\%)}                           & 0.2016 \scriptsize{(+5.3\%)}                             & 0.3576 \scriptsize{(+3.6\%)}                           & 0.1862 \scriptsize{(+6.7\%)}
\\\hline\hline
\end{tabular}
}   
\end{threeparttable}    
\label{tab:synthetic-result}  
\vspace{-0.cm}
\end{table*}

\begin{table*}[t]
\caption{Effects of the noisy control parameter $k$, the best performance for each method and setting are labeled by bold fonts.}
\vspace{-0.3cm}
\center
\small
\renewcommand\arraystretch{1.1}
\setlength{\tabcolsep}{7.1pt}
\begin{threeparttable}  
\scalebox{1.}{
\begin{tabular}{c|c|cc|cc|cc|cc}
\hline\hline
&$N_y$& \multicolumn{4}{c|}{$\mathcal{N}(\mathbf{0,0.2})$}                                             & \multicolumn{4}{c}{0}  \\
\hline
&Dimension     & \multicolumn{2}{c|}{p=16}                                           & \multicolumn{2}{c|}{p=32}
                & \multicolumn{2}{c|}{p=16}
                & \multicolumn{2}{c}{p=32}\\
\hline
&Metrics    &
\multicolumn{1}{l}{{HR@10}} & \multicolumn{1}{l|}{{NDCG@10}} & \multicolumn{1}{l}{{HR@10}} & \multicolumn{1}{l|}{{NDCG@10}} & \multicolumn{1}{l}{{HR@10}} & \multicolumn{1}{l|}{{NDCG@10}} & \multicolumn{1}{l}{{HR@10}} & \multicolumn{1}{l}{{NDCG@10}} \\
\hline

\multirow{3}{*}{CPR-BPR}& $k=1$        & 0.3591                          & 0.1876                            & \textbf{0.3119}                          & \textbf{0.1780}                            & \textbf{0.4024}                          & \textbf{0.2268}                            & 0.3663                          & 0.1971                            \\
&$k=2$        & \textbf{0.3609}                          & \textbf{0.2011}                            & 0.3098                          & 0.1762                            & 0.3936                          & 0.2250                            & \textbf{0.3705}                         & \textbf{0.2028}                            \\
&$k=3$     & 0.3580                          & 0.1891                           & 0.3011                          & 0.1665                            & 0.3877                          & 0.2090                            & 0.3647                          & 0.1914                            \\
\hline
\multirow{3}{*}{CPR-GMF}& $k=1$        & \textbf{0.3437}                          & \textbf{0.1827}                         & \textbf{0.2994}                          &\textbf{0.1593}                            & 0.3725                          & \textbf{0.1983}                            & 0.3470                          & 0.1869                            \\
&$k=2$        & 0.3402                          & 0.1762                            & 0.2941                          & 0.1575                            & \textbf{0.3808}                          & 0.1944                            & 0.3576                         & 0.1891                            \\
&$k=3$       & 0.3354                          & 0.1709                           & 0.2948                          & 0.1588                            & 0.3564                          & 0.1841                            & \textbf{0.3800}                          & \textbf{0.1945}                            \\
\hline
\multirow{3}{*}{CPR-MLP}& $k=1$        & \textbf{0.3580}                          & \textbf{0.1812}                            & \textbf{0.3051}                          & \textbf{0.1560} & 0.3592                 & 0.1868   & 0.3625     & 0.1853                                                       \\
&$k=2$        & 0.3537                          &0.1794                            & 0.2878                          & 0.1465                            & 0.3820                          & 0.1937                            & 0.3611                         & 0.1840                            \\
&$k=3$        & 0.3574                          & 0.1798                           & 0.2886                          & 0.1469                            & \textbf{0.3831}                          & \textbf{0.1940}                            & \textbf{0.3668}                          & \textbf{0.1897}                            \\
\hline
\multirow{3}{*}{CPR-NeuMF}& $k=1$        & \textbf{0.3559}                               & \textbf{0.1800}                              & \textbf{0.3046}                           & 0.1607                              & 0.3865                             & 0.2005                               & 0.3631                             & 0.1851                            \\
&$k=2$        & 0.3511                          & 0.1799                  & 0.2870                          & 0.1412                            & 0.3939                          & 0.2115                            & \textbf{0.3724}                         & \textbf{0.2019}                            \\
&$k=3$        & 0.3552                          & 0.1793                           & 0.2990                          & \textbf{0.1644}                            &\textbf{0.4007}                  &\textbf{0.2166}                            & 0.3643                          & 0.1883                            \\\hline

\multirow{3}{*}{CPR-LightGCN}& $k=1$    & \textbf{0.3512}                               & \textbf{0.1805}                              & 0.2935
                           & 0.1556                              & 
0.3739                             & 0.1952                               & 
\textbf{0.3550}                             & 
\textbf{0.1886}                            \\
&$k=2$        & 
0.3365                          & 0.1764                  & 0.2794                          & 0.1499                            & 
\textbf{0.3818}                          & \textbf{0.2014}                            & 0.3528                         & 
0.1876                            \\
&$k=3$       & 
0.3489                          & 0.1791                           & \textbf{0.2962}
                          & \textbf{0.1569}                            &0.3583                  &0.1874                            & 
0.3462                          & 
0.1813  \\
\hline\hline
\end{tabular}
}   
\end{threeparttable}    
\label{tab:hyperparameter}
\vspace{-0.1cm}
\end{table*}

\subsubsection{Baselines}\label{base}
In order to evaluate the effectiveness of our proposed framework, we compare our model with the following representative methods:
\textbf{ItemPop} is a non-personalized model, and the ranking score of an item is based on its popularity in the training set.
\textbf{ItemKNN}~\cite{ItemKNN_WangVR06} is an item-based k-nearest-neighborhood (KNN) model.
\textbf{BPR}~\cite{rendle2009bpr} is a well-known recommender model based user implicit feedback. 
\textbf{GMF}, \textbf{MLP} and \textbf{NeuMF}~\cite{he2017neural} are neural recommender models, among which NeuMF is a combination between GMF and MLP.
\textbf{CDAE}~\cite{CDAE_WuDZE16} is a method based on denoising auto-encoder, which is proved to be the generalization of several well-known collaborative filtering models.
\textbf{LightGCN}~\cite{LightGCN_0001DWLZ020} is a graph-based recommender model, where the user-item structure information is considered in the modeling process. 

In order to verify the generality of our framework, we apply it to {MF}, {GMF}, {MLP}, {NeuMF} and {LightGCN}, respectively, which leads to the models of \textbf{CPR-MF}, \textbf{CPR-GMF}, \textbf{CPR-MLP}, \textbf{CPR-NeuMF} and \textbf{CPR-LightGCN}.
To demonstrate the necessity of learning-based intervention method, we also evaluate the performance our framework with random intervention, where the recommendation list $\bm{R}$ is assigned in a random manner.
We denote the corresponding baselines as \textbf{CPR-MF-r}, \textbf{CPR-GMF-r}, \textbf{CPR-MLP-r}, \textbf{CPR-NeuMF-r} and \textbf{CPR-LightGCN-r}.

\subsubsection{Implementation details}
We follow the previous work~\cite{rendle2009bpr,he2017neural,he2018adversarial} to employ standard leave-one-out protocol for evaluation.
10 items are recommended from each model to compare with the users' actually interacted ones.
We leverage the widely used metrics including Hit Ratio (HR) and Normalized Discounted Cumulative Gain (NDCG) to compare different models.
Among these metrics, HR aims to measure the overlapping between the predicted and ground truth items, while NDCG further takes the ranking of the prediction results into consideration, and higher ranked accurate items contribute more to the final score.
The implementations of our model and all the baselines are based on PyTorch~\cite{Pytorch_PaszkeGMLBCKLGA19} with mini-batch Adam~\cite{Adam_KingmaB14} optimizer. 
For ItemKNN and CDAE, we follow the settings of the original paper. For the other baselines, we determine their hyper-parameters based on grid search, and the search ranges for the embedding size, batch size, regularization coefficient and learning rate are set as 
$\{32,64,128,256, 512\}$, $\{256, 512, 1024, 2048\}$, $\{2\times 10^{-2},10^{-2},5\times 10^{-3},2\times 10^{-3},10^{-3},5\times 10^{-4}, 10^{-4},10^{-5}\}$ and $\{5\times 10^{-3},2\times 10^{-3}, 10^{-3}\}$, respectively.
The negative samples are selected from the whole item set or the impression list. 
% The negative item number is empirically set as 5.
% For fair comparison, the target models in our CPR framework follow the same settings as the baselines.
When optimizing $p_{\bm{R}}$ and $p_{\bm{S}}$, we empirically set the learning rate as 0.001, and the user/item embedding sizes are both tunned in $\{16,32,64,128,256,512\}$.
The length of the impression list (i.e., $|\bm{R}|$) is set as 5, and the size of $\bm{S}$ (i.e., $k$) is determined in $\{1,2,3\}$.
The Gaussian policy is implemented as a two-layers fully-connected neural network, where the hidden dimension is searched in $\{16, 32, 64\}$.

\subsection{Experiments with Synthetic Dataset}
In this subsection, we present and analyze the experimental results on the synthetic dataset.

\subsubsection{Overall performance}
In the experiments, we explore different user/item representation dimensions as well as both linear and non-linear user response models. 
The overall comparison results are shown in Table~\ref{tab:synthetic-result}.
It is interesting to see that the relative simple baselines (e.g., ItemPop and ItemKNN) are competitive in many cases.
This observation is consistent with the previous work~\cite{recsys_DacremaCJ19}, and suggests that simple methods may still be useful in some recommendation scenarios.

It is encouraging to see that our framework can consistently improve the performance of different target models.
The improvement is consistent across different settings.
In specific, our framework can on average improve the performance of the target model by about {10.08\% and 11.17\%} on HR and NDCG, respectively.
This result demonstrates the effectiveness of our framework.
The reason can be that the recommender simulator enables us to explore the potential user preferences, which provides useful signals to widen the model visions and improve the performance.
The importance of the learning-based intervention method is evidenced in the lowered performance when we use random strategy as a replacement, which verifies our claims in the introduction. 
It is interesting to see that the performance gains on neural models (e.g, MLP, NeuMF and LightGCN) are usually larger than that of the matrix factorization method.
In many cases, the performance is even lowered for GMF and BPR after applying our framework, e.g., when we use non-linear user response model and set $d=32$.
This observation agrees with our expectation, that is, neural models need more training samples to display its strong expressiveness, and our framework should be more effective for them.
Since most of the promising recommender models in practice are based on neural architectures, this observation demonstrates the potential of our framework in real-world settings.

\begin{table}[t]
\caption{{Overall results on the real-world dataset.}}
\vspace{-0.2cm}
\center
\small
\renewcommand\arraystretch{1.1}
\setlength{\tabcolsep}{10.pt}
\scalebox{1.0}{
\begin{tabular}{ccc}
\hline\hline
Model           & H@10 & NDCG@10 \\
\hline
ItemPop         & 0.3374     & 0.1608        \\
ItemKNN         & 0.2403     & 0.1299        \\
CDAE            & 0.3403     & 0.1637        \\
\hline
BPR             & 0.3440     & 0.1685        \\
CPR-BPR-r       & 0.3524     & 0.1718        \\
CPR-BPR         & 0.3526 \scriptsize{(+2.50\%)}     & 0.1810 \scriptsize{(+7.41\%)}        \\
\hline
GMF             & 0.3453     & 0.1651        \\
CPR-GMF-r       & 0.3525     & 0.1686        \\
CPR-GMF         & 0.3519 \scriptsize{(+1.91\%)}     & 0.1713 \scriptsize{(+3.75\%)}        \\
\hline
MLP             & 0.3469     & 0.1662        \\
CPR-MLP-r       & 0.3560     & 0.1702        \\
CPR-MLP         & 0.3526 \scriptsize{(+1.64\%)}     & 0.1747 \scriptsize{(+5.11\%)}        \\
\hline
NeuMF           & 0.3381     & 0.1630        \\
CPR-NeuMF-r     & 0.3466     & 0.1662        \\
CPR-NeuMF       & 0.3512 \scriptsize{(+3.87\%)}     & 0.1755 \scriptsize{(+7.66\%)}        \\
\hline
LightGCN        & 0.3451     & 0.1658        \\
CPR-LightGCN-r  & 0.3521     & 0.1719        \\
CPR-LightGCN    & 0.3688 \scriptsize{(+6.86\%)}     & 0.1830 \scriptsize{(+10.4\%)}        \\
\hline\hline
\end{tabular} 
}       
\label{tab:mind-result}   
\vspace{-0.cm}
\end{table}

\begin{table*}[t]
\caption{{Performances on the items with different coldnesses.}}
\vspace{-0.cm}
\small
\begin{tabular}{c|cc|cc|cc}
\hline\hline
Popularity   & \multicolumn{2}{c|}{Low}                                             & \multicolumn{2}{c|}{Middle}                                          & \multicolumn{2}{c}{High}                                            \\\hline
Metrics      & HR@10 & NDCG@10 & HR@10 & NDCG@10 & HR@10 & NDCG@10 \\
\hline
BPR           & 0.0367   & 0.0113  & 0.4165  & 0.1426 & 0.8966  & 0.4937                            \\
CPR-BPR      & 0.0380 \scriptsize{(3.54\%)}   & 0.0121 \scriptsize{(7.08\%)}  & 0.4494 \scriptsize{(7.90\%)}  & 0.1602 \scriptsize{(12.34\%)}  & 0.8930 \scriptsize{(-0.40\%)} & 0.5027 \scriptsize{(1.82\%)}  \\\hline

GMF        & 0.0079    & 0.0023     & 0.3715    & 0.1163    & 0.9157   & 0.4793   \\
CPR-GMF    & 0.0092 \scriptsize{(16.46\%)}  & 0.0027 \scriptsize{(17.39\%)}   & 0.4028 \scriptsize{(8.43\%)}  & 0.1270 \scriptsize{(9.20\%)}   & 0.9152 \scriptsize{(-0.05\%)}& 0.4810 \scriptsize{(0.35\%)} \\
\hline

MLP          & 0.0154 & 0.0046 & 0.3531  & 0.1130  & 0.9240  & 0.4845   \\
CPR-MLP      & 0.0196 \scriptsize{(27.27\%)}  & 0.0059 \scriptsize{(28.26\%)}  & 0.4091 \scriptsize{(15.86\%)}  & 0.13656 \scriptsize{(20.85\%)} & 0.9261 \scriptsize{(0.23\%)} & 0.4870 \scriptsize{(0.52\%)} \\\hline

NeuMF        & 0.0246 & 0.0075 & 0.3010 & 0.0970  & 0.9018   & 0.4782                            \\
CPR-NeuMF    & 0.0277 \scriptsize{(12.60\%)}  & 0.0085 \scriptsize{(13.33\%)}  & 0.3940 \scriptsize{(30.90\%)} & 0.1271 \scriptsize{(31.03\%)} & 0.9182 \scriptsize{(1.82\%)} & 0.4872\scriptsize{(1.88\%)} \\\hline

LightGCN     & 0.0172& 0.0051& 0.3385& 0.1061& 0.9171& 0.4781\\
CPR-LightGCN & 0.0433 \scriptsize{(151.74\%)} & 0.0134 \scriptsize{(162.75\%)}  & 0.4769 \scriptsize{(40.89\%)} & 0.1601 \scriptsize{(50.90\%)}  & 0.9115 \scriptsize{(-0.61\%)}                          & 0.4860 \scriptsize{(1.65\%)}                           \\\hline\hline
\end{tabular}
\label{tab:pop-result}   
\vspace{-0.cm}
\end{table*}

\subsubsection{Effects of the noisy control parameter $k$}
In this section, we investigate the effectiveness of the noisy control method proposed in section~\ref{theory}.
By this method, we hope our framework can adaptively tune itself to accommodate different noisy recommendation scenarios.
In order to evaluate such capability, we set $N_y=0$ or sample it from $\mathcal{N}(\mathbf{0,0.2})$ to simulate the settings where the data is observed with different noisy levels.
We base the experiment on the non-linear user response model, and we tune $k$ in the range of $\{1,2,3\}$.
The results are presented in table~\ref{tab:hyperparameter}. 
It is not surprising to see that the performance of the same model is lowered when the noisy level is increased from $0$ to sampling from $\mathcal{N}(\mathbf{0,0.2})$.
When $N_y$ is sampled from $\mathcal{N}(\mathbf{0,0.2})$, the best performance is mostly achieved when $k=1$, while if we set $N_y=0$, $k=2$ or $k=3$ can usually lead to more favored results.
We speculate that: when the dataset is noisy, the recommender simulator can be not trained well, thus the quality of the generated samples are not high.
In such a scenario, removing the noisy samples seems to be more important.
On the contrary, if the dataset is noise-free, then the recommender simulator can be more accurate. 
In this case, the sample quality is not the main issue, and more samples are favored to achieve sufficient model optimization and better performance.
For different noisy-level datasets, our framework can always adapt itself to achieve better performance, which demonstrates the effectiveness of the noisy control method.

% \begin{figure*}[t]
% \centering
% \setlength{\fboxrule}{0.pt}
% \setlength{\fboxsep}{0.pt}
% \fbox{
% \includegraphics[width=.7\linewidth]{fig/real.png}
% }
% \caption{(a) Performance comparison based on the real-world dataset. }
% \label{real-world}
% \end{figure*}

\subsection{Experiments with Real-World Dataset}
The above synthetic experiments suggest that our idea is effective under ideal environment. 
In this section, we experiment with the real-world dataset, which is more challenging but more practical.

\subsubsection{Overall comparison}
To begin with, we compare our model with the baselines introduced in section~\ref{base}.
The results are presented in Table~\ref{tab:mind-result}, from which we can see: similar to the synthetic dataset, our CPR framework can consistently lead to improved performances comparing with the target models, and the learning-based intervention method outperforms the random strategy in most cases.
The performance gains on neural models are usually larger than that of the shallow ones, which again demonstrates the potential of our model for deep recommender models.
Remarkably, the improvement of our framework is not that large comparing with the results on the synthetic dataset. We speculate that the real-world dataset can be noisy. Training based on it can lead to imperfect recommender simulator, which lowers the quality of the generated samples and limits the final recommendation performance.

\subsubsection{Performance on the items with different coldness}
Cold start has long been a fundamental problem in the recommendation domain.
In this section, we study the effectiveness of our framework on the items with different coldnesses.
More specifically, we first run the models based on the original training set, and then evaluate their performances on three different testing sets varying on the item coldness.
For the first testing set (denoted as ``Low''), each item has less than 5 interactions with the users in the training set, which means the model is trained in a quite insufficient manner on these items.
In the second testing set (denoted as ``Middle''), each item has 5 to 15 interactions, thus the model is provided with more knowledge during the training process.
In the last testing set (denoted as ``High''), each item has more than 15 interactions, which is the ``warmest'' setting in the experiment.
The model parameters are set as their optimal values tunned in the above experiments. 

From Table~\ref{tab:pop-result} we can see: our framework can enhance the performance of the target models in most cases, which demonstrates its effectiveness under different cold start settings. An interesting observation is that the performance improvement on the ``Low'' and ``Middle'' testing sets are much larger than that of the ``High'' dataset. 
% For example, when applying CPR on MLP, the improvement on the ``Low'' and ``Middle'' datasets are about 15\%-30\%, while on the ``High'' dataset, the improvement is less than 1\%.
For example, CPR-LightGCN can produce a remarkable 151.74\% and 162.75\% improvements over LightGCN on HR@10 and NDCG@10 for the ``Low'' dataset, and the improvements on the ``Middle'' dataset is about 40\%-50\%. However, on the ``High'' dataset, the performance is even lowered on the metric of HR@10, and the improvement on NDCG@10 is also limited.
We speculate that, in the ``Low'' dataset, the testing items are trained in a quite insufficient manner.
A large amount of knowledge has been ignored by the target model, which provides more opportunities for the generated samples to introduce useful signals and improve the performance.
In real-world recommendation scenarios, cold-start is a notorious problem obsessing people for a long time, this experiment demonstrates the potential of our framework in alleviating this problem.

\vspace{-0.1cm}
\section{Related Work}
In this section, we compare our model with the previous work to highlight the contributions of this paper.

\noindent
\textbf{Relation with causal recommendation.}
Recent years have witnessed many contributions on incorporating causal inference into the recommendation domain~\cite{dong2020counterfactual,chen2020bias,yang2018unbiased,saito2019unbiased,schnabel2016recommendations}.
For example, ~\cite{schnabel2016recommendations} explains the recommendation problem by a treatment-effect model, and designs a re-weighting method to remove the bias in the observed data. However, this method is based on user explicit feedback, and the loss function is restricted to root mean square error (RMSE).
In order to handle user implicit feedback, ~\cite{yang2018unbiased,saito2019unbiased} extend this method by incorporating cross-entropy loss and designing tailored debiasing models.
In addition, ~\cite{liu2020general} proposes a general knowledge distillation framework to debias the training data.
~\cite{chen2020bias} provides a thorough discussion on the recent progress on debiased recommendation.
~\cite{bonner2018causal} leverages uniform data to learn causal user/item embeddings for more fair and unbiased recommendation.
These methods have achieved many successes in the recommendation domain. However, they mostly leverage causal inference to debias the training data, which is different from our data augmentation purpose. In addition, previous work mostly follow Rubin's potential outcome framework.
However, our idea is inspired from Pearl's structure causal models, where we can explicitly model the data generated environment to ensure compatibleness between the generated and observed data.

\noindent
\textbf{Relation with counterfactual data augmentation.}
Counterfactual data augmentation stems from the human introspection behavior.
It has been recently leveraged to alleviate the training data insufficiency problem in the machine learning community.
In the past few years, this idea has achieved great successes in the fields of neural language processing (NLP)~\cite{zmigrod2019counterfactual} and computer vision (CV)~\cite{fu2020counterfactual,chen2019counterfactual,ashual2019specifying}.
In this paper, we apply it to the top-N recommendation task, which, to the best of our knowledge, is the first time in this field.
The major differences between our framework and the previous work is: (\romannumeral1) we design a leaning-based intervention method to encourage the informativeness of the generated samples.
(\romannumeral2) We theoretically analyze the noisy information in the generated samples, and design a tailored noise control method.

\noindent
\textbf{Relation with ranking based recommendation model.}
Ranking based recommender models have been widely studied for the Top-N recommendation task.
The most famous model in early years is Bayesian personalized ranking (BPR), which is optimized by maximizing the user preference margin between the positive and negative items.
This method has inspired many promising models.
For example, CKE~\cite{KDD_ZhangYLXM16} proposes a hybrid model to integrate collaborative filtering and knowledge base for recommendation. 
~\cite{ding2018improved} improves BPR with a better negative sampler which leverages additional data in E-commerce. 
AMF~\cite{he2018adversarial} applies adversarial training method to enhance the performance of BPR. 
With the ever prospering of deep neural network, recent years have witnessed the surge of neural recommender models~\cite{he2017neural,cheng2016wide,guo2018deepfm,covington2016deep,wang2017deep}.
For example, NeuMF~\cite{he2017neural} is designed to capture the non-linear user-item correlations.
NGCF~\cite{wangxiang1} and LightGCN~\cite{LightGCN_0001DWLZ020} regard the user-item interactions as a graph, and explicitly incorporate the structure information to enhance the utilization of the collaborative filtering signals.
% leverage graph neural network to propagate more comprehensive collaborative filtering signals, which can help to enhance the user/item representations and improve the final recommendation performance.
By looking back the history of ranking based recommendation models, it is evident that the model architectures are becoming deeper and heavier.
While the comprehensive parameters can indeed lead to improved recommendation performance, an unprecedented problem is also emerging, that is, more training data is needed to satisfy the heavy architectures, which contradicts with the sparse user behaviors in realities.
In this paper, we propose a solution to this problem based on causal inference, which is parallel with the previous model-drive research.

\vspace{-0.2cm}
\section{Conclusion}
In this paper, we propose to enhance Top-N recommendation with Pearl's causal inference framework, where we can simulate user preference and generate counterfactual samples for alleviating the training data insufficiency problem.
We design a learning-based intervention method for discovering the informative samples, and conduct theoretical analysis to reveal the relation between the number of generated samples and the potential model prediction error. Extensive experiments based on both synthetic and real-world datasets are conducted to verify our model's effectiveness. 

This paper actually opens the door of incorporating Pearl's causal inference framework into the recommendation domain. There is still much room for improvement. For example, one can design more advanced exogenous node structures to introduce more reasonable prior knowledge for different recommendation scenarios.
In addition, it should be also interesting to incorporate side information into the structure equation models, which can help to capture more comprehensive user preference and obtain more accurate recommender simulators.

\section{Appendix}
\subsection{Proof of the Theories in Section~\ref{theory}}

\subsubsection{Proof of theory~\ref{t}}
\begin{proof}
Without loss of generality, we suppose $\eta > \frac{1}{2}$.
We define the random variable $X_i$ as the $i$th observation of the relation between $i$ and $j$.
We use $X_i = 1$ to represent $i>j$, and $X_i = 0$ to indicate $i<j$.
Suppose, we have N observations, and we define $\overline{X} = \frac{1}{N}\sum_{i=1}^N X_i$.
Following the voting mechanism, if $\overline{X}< \frac{1}{2}$, then the prediction is wrong, since it contradicts with $\eta > \frac{1}{2}$.
Obviously,
\begin{equation}
E[\overline{X}] = \frac{1}{N}\sum_{i=1}^N E[X_i] = \eta.
\end{equation}
According to the Hoeffding's inequality~\cite{shalev2014understanding}, we have:
\begin{equation}
p(\overline{X}<\frac{1}{2}) =  p(\overline{X}-E[\overline{X}]<\frac{1}{2}-\eta) < \exp{(-2N(\frac{1}{2}-\eta)^2)}.
\end{equation}
By setting $\delta = \exp{(-2N(\frac{1}{2}-\eta)^2)}$, we have when $N$ is larger than $\frac{\log{\frac{1}{\delta}}}{2(1-2\eta)^2}$, $p(\overline{X}<\frac{1}{2})< \delta$, which means the prediction error is smaller than $\delta$. 
\end{proof}

\subsubsection{Proof of theory~\ref{t1}}
\begin{proof}
For a hypothesis $f\in \mathcal{F}$, suppose its prediction error is $s$ (i.e., $\sum \mathbb{I}(g_f \neq g) = s$)\footnote{$g_f$ is the prediction from $f$, and $g$ is the ground truth.}, then the mis-matching probability between the observed and predicted results comes from two parts:

$\bullet$ The observed result is true, but the prediction is wrong, that is, $s(1-\zeta)$.

$\bullet$ The observed result is wrong, but the prediction is right, that is $(1-s)\zeta$.

Thus, the total mis-matching probability is $\zeta + s(1-2\zeta)$.
Suppose the prediction error of $f$ (i.e., $s$) is larger than $\epsilon$, 
Then, at least one of the following statements hold:

$\bullet$ The empirical mis-matching rate of $f$ is smaller than $\zeta+\frac{a(1-2\zeta)}{2}$.

$\bullet$ The empirical mis-matching rate of the optimal $f^*$ is larger than $\zeta+\frac{\epsilon(1-2\zeta)}{2}$.

These statements are easy to understand, since if both of them do not hold, we can conclude that the empirical loss of $f$ is larger than that of $f^*$, which does not agree with the ERM definition.
However, according to the Hoffding inequality~\cite{shalev2014understanding}, both of the above statements hold with the probability smaller than $\delta$, which implies that the prediction error of $f$ is smaller than $\epsilon$ with the probability larger than $1-\delta$.
\end{proof}

\bibliographystyle{ACM-Reference-Format}
\balance
\bibliography{acmart}

\end{document}